\documentclass[10pt,twocolumn,aps,pra,amsmath,amssymb,showpacs]{revtex4-1}
\usepackage{bm}
\usepackage{mathrsfs}
\usepackage{graphicx}
\usepackage[usenames,dvipsnames,svgnames,table]{xcolor}
\usepackage[unicode=true,
            pdfusetitle, 
            bookmarks=true,
            bookmarksnumbered=false,
            bookmarksopen=false,
            breaklinks=true,
            pdfborder={0 0 0},
            backref=false,
            colorlinks=true]{hyperref}
\hypersetup{linkcolor=NavyBlue,urlcolor=NavyBlue,citecolor=NavyBlue}

\usepackage{amsthm}
\newtheorem{prop}{\protect\propositionname}
\providecommand{\propositionname}{Proposition}
\newtheorem{thm}{Theorem}

 \newtheorem{cor} {Corollary}
 \newtheorem{rem}{Remark}

 \usepackage{tikz-cd}

\newcommand{\cH}{\mathcal{H}}

\begin{document}

\title {Positivity of the assignment map  implies complete positivity of the reduced dynamics}

\author {Iman Sargolzahi}
\email{sargolzahi@neyshabur.ac.ir; sargolzahi@gmail.com}
\affiliation {Department of Physics, University of Neyshabur, Neyshabur, Iran}

\affiliation{Research Department of Astronomy and Cosmology, University of Neyshabur, Neyshabur, Iran}

\begin{abstract}
Consider the set $\mathcal{S}=\lbrace\rho_{SE}\rbrace$ of possible initial states of the system-environment. The map which assigns to each $\rho_{S}\in \mathrm{Tr}_{E}\mathcal{S}$ a $\rho_{SE}\in \mathcal{S}$ is called the assignment map. The assignment map is  Hermitian, in general. In this paper, we restrict ourselves to the case that the assignment map is, in addition, positive and 
 show that this implies that the so-called \textit{reference state} is a Markov state. Markovianity of the reference state leads to existence of another assignment map which is completely positive.
  So, the reduced dynamics of the system is also  completely positive.
  As a consequence, when the system $S$ is a qubit, we show that if  $\mathcal{S}$ includes entangled states, then either the reduced dynamics  is not given by a map, for, at least, one unitary   time evolution of the system-environment $U$, or the reduced dynamics is non-positive, for, at least, one $U$.
\end{abstract}

\maketitle
\section{Introduction}

In quantum information theory, it is common to assume that quantum operations are linear trace-preserving completely positive (CP) maps   \cite{3}:
\begin{equation}
\label{eq:a3}
\begin{aligned}
\rho^\prime=\Phi(\rho)=\sum_{i}E_{i}\, \rho \,E_{i}^{\dagger},\ \ \ \sum_{i}E_{i}^{\dagger}E_{i}=I,
\end{aligned}
\end{equation}
where $\rho$ and $\rho^\prime$ are the initial  and final states (density operators) of the system, respectively. In addition, $E_{i}$ are linear operators, and $I$ is the identity operator.

 For example, consider a bipartite quantum system $S=AB$. In entanglement theory, it is assumed that the entanglement, between the two separated parts $A$ and $B$, does not increase under local quantum operations  \cite{25-a}. Assuming quantum operations as CP maps, an entanglement measure (monotone) $\mathcal{M}$ is constructed as a non-increasing function, under local CP maps \cite{25-a, a17}, i.e.,  
\begin{equation}
\label{eq:aa3}
\begin{aligned}
\mathcal{M}\left(\rho_{AB}\right) \geq \mathcal{M} \left( \Phi_A \otimes \Phi_B (\rho_{AB})\right),
\end{aligned}
\end{equation} 
where $\Phi_A$ and $\Phi_B$ are CP maps, as Eq. \eqref{eq:a3}, on the parts $A$ and $B$, respectively. 

Now, consider the case that
 the bipartite system $S=AB$ is not closed and interacts with its environment $E=E_AE_B$, where $E_A$ and $E_B$ are the local environments of the parts $A$ and $B$, respectively.
 In addition, assume that the time evolution of the whole system-environment is local as $U_{SE}=U_{AE_A}\otimes U_{BE_B}$, where $U_{AE_A}$ and  $U_{BE_B}$ are unitary operators on $AE_A$ and $BE_B$, respectively. Then, the reduced dynamics of the system $S=AB$  may not be given as a local CP map $\Phi_A \otimes \Phi_B$,  in general, and so the entanglement may increase, during such a local evolution of the whole system-environment \cite{a8, a2b, a6, a7, a5, a11a, a11}.
 
 In entanglement theory, one usually consider only one initial state $\rho_{AB}$, and study the entanglement evolution of the system, from this  initial state $\rho_{AB}$ to the final state  $\rho_{AB}^\prime$. It can be shown that 
  the evolution  from  \textit{one}  state $\rho_{AB}$ to another state  $\rho_{AB}^\prime$ can always  be written as a CP map $\Phi_S=\Phi_{AB}$ \cite{9}. But, this $\Phi_S$ is not local, in general: $\Phi_S \neq \Phi_A \otimes \Phi_B$. So, the entanglement may increase, during the evolution. The circumstances, for which  the reduced dynamics of the system $S=AB$ can be given as a local CP map $\Phi_A \otimes \Phi_B$, have been studied in \cite{a12}. 

In contrast to the entanglement theory, one may find important theorems, in quantum information theory, in which the evolution of more than one initial state is considered, e.g., the evolution of the \textit{trace distance} \cite{3}, between two states $\rho$ and $\sigma$. Since there is more than one initial state, we cannot use the result of \cite{9}. Therefore, the time evolution of the system $S$ may be non-CP, in general, which can lead to unexpected results.

For example, it is known that the trace distance does not increase, under CP maps \cite{3}. Assuming that the most general quantum operation is CP results that the  trace distance is always contractive. But, in an open quantum system, it is possible to find a Hermitian non-positive reduced dynamics for which the trace distance, between two states, increases, after the evolution \cite{a14}. There exists a similar result for the\textit{ relative entropy} \cite{3} of two states $\rho$ and $\sigma$. The relative entropy does not increase, under CP maps \cite{3, a15, a16}, and even under positive maps \cite{12}. But, one can find Hermitian non-positive reduced dynamics for which the relative entropy increases, after the evolution \cite{a14, 25}.

The above examples show that in open quantum systems, at least, when we deal with more than one initial state, it is important to determine whether the reduced dynamics of the system is CP or not.

Reduced dynamics of a quantum system $S$, interacting with its environment $E$, is given by 
\begin{equation}
\label{eq:1.1}
\begin{aligned}
\rho_{S}^{\prime}=\mathrm{Tr}_{E} \circ \mathrm{Ad}_{U}(\rho_{SE})\equiv \mathrm{Tr}_{E}\left( U \rho_{SE}U^{\dagger}\right), 
\end{aligned}
\end{equation} 
where $\rho_{SE}$ is the initial state of the system-environment  and the unitary operator $U=U_{SE}$ acts on the whole Hilbert space of the system-environment. The initial state of the system is $\rho_{S}=\mathrm{Tr}_{E} (\rho_{SE})$.
Now, if $\rho_{SE}$ can be written as a function of   $\rho_{S}$, i.e., 
\begin{equation}
\label{eq:1.2}
\begin{aligned}
\rho_{SE}= \Lambda_S (\rho_{S}),
\end{aligned}
\end{equation} 
where $\Lambda_S$ is called the \textit{assignment map} \cite{1,2}, then the final state of the system is given by the following map
\begin{equation}
\label{eq:1.3}
\begin{aligned}
\rho_{S}^{\prime}=\mathrm{Tr}_{E} \circ \mathrm{Ad}_{U}  \circ \Lambda_S(\rho_{S})\equiv 
\mathcal{E}_S(\rho_{S}). 
\end{aligned}
\end{equation} 
The unitary evolution $U$ and the partial trace $\mathrm{Tr}_{E}$ are CP maps \cite{3}. The assignment map $\Lambda_S$ is, in general, Hermitian \cite{4}, i.e., maps each Hermitian operator to a Hermitian operator. 
Therefore, 
   the \textit{dynamical map} $\mathcal{E}_S$ is, in general, Hermitian \cite{5, 5-a}.
  

It was known that if the assignment map $\Lambda_S$ is (a)   positive, i.e., maps each positive operator to a positive operator, and (b) consistent, i.e.,  $\mathrm{Tr}_{E}(\Lambda_S(\rho_{S}))=\rho_{S}$, for all $\rho_{S}$, on the Hilbert space of the system  $\mathcal{H}_S$, then it is given by
\begin{equation}
\label{eq:1.4}
\begin{aligned}
\Lambda_S(\rho_{S})=\rho_{S}\otimes\sigma_E,
\end{aligned}
\end{equation} 
where $\sigma_E$ is a fixed state, on the Hilbert space of the environment  $\mathcal{H}_E$ \cite{1, 6}.

Interestingly, the above \textit{Pechukas's assignment map} is not only positive, but also CP, and so the reduced dynamics $\mathcal{E}_S$, in Eq. \eqref{eq:1.3}, is also CP.
Whether this result can be generalized to arbitrary positive assignment map, is the subject of this paper.

We consider the case that there exists a linear positive assignment map $\Lambda_S$, but we relax the condition (b) of the consistency of $\Lambda_S$ for arbitrary $\rho_S$.
Then, we show that the positivity of   $\Lambda_S$ implies that the \textit{reference state}, introduced in \cite{7}, is a so-called \textit{Markov state}, as defined in \cite{8}.

Markovianity of the reference state results in existence of another assignment map $\Lambda_S^{(CP)}$, which is CP, and the CP-ness of  $\Lambda_S^{(CP)}$ leads to the CP-ness of the reduced dynamical map 
$\mathcal{E}_S$, in Eq. \eqref{eq:1.3}.

The above result, as our main result, will be given in Sec. \ref{sec: C}.
Before, in Sec. \ref{sec: B}, we review the preliminaries needed to achieve the above result, including the reference state and the role of its (non-)Markovianity on the (non-)CP-ness of the assignment map.
 
In Sec.  \ref{sec: C}, we consider the case that there exists a one to one correspondence between the members of the set $\mathcal{S}=\lbrace\rho_{SE}\rbrace$, of possible initial states of the system-environment, and the members of the set  $\mathcal{S}_S\equiv\mathrm{Tr}_{E}\mathcal{S}$. Generalization to the arbitrary case will be given in Sec.  \ref{sec: D}.

As an application of our result, we consider the case that the system $S$ is a qubit, in Sec.  \ref{sec: E}.
 We show that when there exists entanglement between the system  and the environment,  
   then either the reduced dynamics of the system  is not given by a map, for, at least, one  $U$, or the reduced dynamics is non-positive, for, at least, one $U$.
 
 Finally, we conclude our paper in Sec.  \ref{sec: F}.

\section{Reduced dynamics and reference state} \label{sec: B}

In this paper, we consider the case that both the system $S$ and the environment $E$ are finite dimensional.
The dimensions of  $\mathcal{H}_S$ and $\mathcal{H}_E$ are
  $d_S$ and   $d_E$, respectively.
We denote the vector space of linear operators $A:\mathcal{H}\rightarrow \mathcal{H}$ 
 by $\mathcal{L}(\mathcal{H})$, and the set of density operators on the Hilbert space  $\mathcal{H}$ by $\mathcal{D}(\mathcal{H})$.

In addition,   we call  a linear trace-preserving Hermiticity-preserving map, simply, a Hermitian map. We denote  a linear trace-preserving positivity-preserving map  as a positive map, and a linear trace-preserving completely
 positive map as a completely positive (CP) map. 
 
 For  each  Hermitian map $\Phi$, on the whole $\mathcal{L}(\cH)$,  there exists an operator sum representation  such that for each $x\in \mathcal{L}(\cH)$, we have
\begin{equation}
\label{eq:10-1}
\begin{aligned}
\Phi(x)=\sum_{i}e_{i}\,\tilde{E_{i}}\,x\,\tilde{E_{i}}^{\dagger},\ \ \ \sum_{i}e_{i}\,\tilde{E_{i}}^{\dagger}\tilde{E_{i}}=I,
\end{aligned}
\end{equation}
where $\tilde{E_{i}}$ are linear operators on $\cH$, and $e_{i}$ are real coefficients \cite{4, 5, 6}. 
When all of the coefficients $e_{i}$ in Eq. \eqref{eq:10-1} are positive,  we  define $E_{i}=\sqrt{e_{i}}\,\tilde{E_{i}}$ and the map $\Phi$ is  CP, as Eq. \eqref{eq:a3}.

\subsection{Reduced dynamics of an open quantum system}  \label{sec: B.1}

Consider the set $\mathcal{S}=\lbrace\rho_{SE}\rbrace$ of possible initial states of the system-environment. Since both the system and the environment are finite dimensional, a finite number $m$ of the members of $\mathcal{S}$, where the integer $m$ is  $0< m\leq {(d_S)}^2{(d_E)}^2$, are linearly independent. Let us denote this linearly independent set as
  $\mathcal{S}^\prime  = \lbrace\rho_{SE}^{(1)}, \rho_{SE}^{(2)}
  , \ldots\ , \rho_{SE}^{(m)}\rbrace$. Therefore, any $\rho_{SE}\in \mathcal{S}$ can be written as $\rho_{SE}=\sum_{i=1}^m a_i \rho_{SE}^{(i)}$, where $a_i$ are real coefficients.

In the following, we restrict ourselves to the case that all $\rho_{S}^{(i)}=\mathrm{Tr}_E(\rho_{SE}^{(i)}) \in \mathcal{S}^\prime_S\equiv \mathrm{Tr}_E\mathcal{S}^\prime$, $i=1, \dots ,  m\leq {(d_S)}^2$, are also linearly independent.
Therefore, there is a one to one correspondence between the members of $\mathcal{S}$ and the members of  $\mathcal{S}_S=\mathrm{Tr}_E\mathcal{S}$. Generalization to the case, that there is no such correspondence, will be given in Sec. ~\ref{sec: D}.

Now, we define the subspace $\mathcal{V}$ as the subspace spanned by $\rho_{SE}^{(i)} \in \mathcal{S}^\prime$ \cite{4}:
\begin{equation}
\label{eq:2.1}
\begin{aligned}
\mathcal{V}= \mathrm{Span_\mathbb{C}}\mathcal{S}^\prime = \mathrm{Span_\mathbb{C}}\mathcal{S}
\subseteq \mathcal{L}(\mathcal{H}_S\otimes\mathcal{H}_E).
\end{aligned}
\end{equation} 
  Therefore, each  $X\in \mathcal{V}$ can be expanded as $X= \sum_{i=1}^m c_i \rho_{SE}^{(i)}$, with complex coefficients $c_i$. In addition, for each $x=\mathrm{Tr}_E(X) \in \mathcal{V}_S \equiv\mathrm{Tr}_E\mathcal{V}$, we have $x= \sum_{i=1}^m c_i \rho_{S}^{(i)}$.
Note that since $\mathcal{S}\subseteq \mathcal{D}(\mathcal{H}_S\otimes \mathcal{H}_E) \cap \mathcal{V}\subset \mathcal{V}$ and $\mathcal{S}_S\subseteq \mathrm{Tr}_E( \mathcal{D}(\mathcal{H}_S\otimes \mathcal{H}_E) \cap \mathcal{V}) \subseteq \mathcal{D}(\mathcal{H}_S)\cap \mathcal{V}_S \subset \mathcal{V}_S$, what which we show for the whole $\mathcal{V}$ and $\mathcal{V}_S$, is also valid for their subsets $\mathcal{S}$ and $\mathcal{S}_S$, respectively.

Since all  $\rho_{S}^{(i)}\in\mathcal{S}_S^\prime$ are linearly independent, as all $\rho_{SE}^{(i)}\in\mathcal{S}^\prime$, for each $x\in \mathcal{V}_S$, there is only one  $X\in \mathcal{V}$ such that $\mathrm{Tr}_E(X)=x$. This allows us to define the 
 linear assignment map $\Lambda_S$ as bellow. We define  $\Lambda_S(\rho_{S}^{(i)})=\rho_{SE}^{(i)}$, $i=1, \ldots, m$. So, for any $x=\sum_{i=1}^m c_i \rho_{S}^{(i)} \in \mathcal{V}_S$, we have
\begin{equation}
\label{eq:2.2}
\begin{aligned}
\Lambda_S(x)=\sum_{i=1}^m c_i \Lambda_S(\rho_{S}^{(i)})=\sum_{i=1}^m c_i \rho_{SE}^{(i)}=X.
\end{aligned}
\end{equation} 

$\Lambda_S$ is a map on the whole $\mathcal{V}_S$. If $m= {(d_S)}^2$, then $\mathcal{V}_S= \mathcal{L}(\mathcal{H}_S)$. Even if $m< {(d_S)}^2$, we can simply generalize $\Lambda_S$ to the  whole $\mathcal{L}(\mathcal{H}_S)$. Consider the set  $\hat{\mathcal{S}}^\prime_S 
 = \lbrace\rho_{S}^{(m+1)},  \ldots\ , \rho_{S}^{({(d_S)}^2)}\rbrace$ such that $\tilde{\mathcal{S}}^\prime_S= \hat{\mathcal{S}}^\prime_S\cup \mathcal{S}^\prime_S$ is a linearly independent set.
So, each $y\in\mathcal{L}(\mathcal{H}_S)$ can be expanded as
  $y=\sum_{i=1}^{{(d_S)}^2} b_i \rho_{S}^{(i)}$, with complex  coefficients $b_i$ and $\rho_{S}^{(i)}\in \tilde{\mathcal{S}}^\prime_S$.
Defining, for $i=m+1, \ldots , (d_S)^2$,  $\Lambda_S(\rho_{S}^{(i)})=\rho_{SE}^{(i)}$, where $\rho_{SE}^{(i)}\in \mathcal{D}(\mathcal{H}_S\otimes \mathcal{H}_E)$ are chosen arbitrarily, we can generalize the assignment map $\Lambda_S$ in Eq. \eqref{eq:2.2} to the whole  $\mathcal{L}(\mathcal{H}_S)$:
\begin{equation}
\label{eq:2.3}
\begin{aligned}
\Lambda_S(y)=\sum_{i=1}^{{(d_S)}^2} b_i  \Lambda_S(\rho_{S}^{(i)})=\sum_{i=1}^{{(d_S)}^2} b_i  \rho_{SE}^{(i)}\equiv Y.
\end{aligned}
\end{equation} 
So, $\Lambda_S: \mathcal{L}(\mathcal{H}_S)\rightarrow\mathcal{L}(\mathcal{H}_S\otimes\mathcal{H}_E)$
 is a Hermitian map, by construction. 
Note that though, for any $x\in \mathcal{V}_S$,  $\Lambda_S(x)=X$  means that $\mathrm{Tr}_E(X)=x$, but for a $y\notin\mathcal{V}_S$, $\Lambda_S(y)=Y$ may lead to $\mathrm{Tr}_E(Y)\neq y$. This is so since we have chosen $\rho_{SE}^{(i)}=\Lambda_S(\rho_{S}^{(i)})$, for  $i=m+1, \ldots , (d_S)^2$, arbitrarily.
In other words, $\Lambda_S$ is consistent only on $\mathcal{V}_S$, and not necessarily on the whole $\mathcal{L}(\mathcal{H}_S)$. 

It is also worth noting that for Hermiticity of $\Lambda_S$, it is enough to choose 
$\Lambda_S(\rho_{S}^{(i)})=B_{SE}^{(i)}$, $i=m+1, \ldots , (d_S)^2$, where $B_{SE}^{(i)}$ are arbitrary 
Hermitian  operators on $\mathcal{H}_S\otimes\mathcal{H}_E$, with unit trace. But, since, in this paper, we are interested in positive assignment maps, we have defined $\Lambda_S(\rho_{S}^{(i)})=\rho_{SE}^{(i)}\in \mathcal{D}(\mathcal{H}_S\otimes \mathcal{H}_E)$, $i=m+1, \ldots , (d_S)^2$.

Now, for any $\rho_S \in  \mathrm{Tr}_E( \mathcal{D}(\mathcal{H}_S\otimes \mathcal{H}_E) \cap \mathcal{V})$, and any unitary evolution $U$ of the whole system-environment, the reduced dynamics of the system is given by Eq. \eqref{eq:1.3}, where the dynamical map $\mathcal{E}_S$ is a Hermitian map on $\mathcal{L}(\mathcal{H}_S)$,  as Eq. \eqref{eq:10-1}.

\subsection{ Reference state}  \label{subsec: B.2}

In Ref. \cite{7}, we have introduced the \textit{reference state} $\omega_{RS} \in  \mathcal{D}(\mathcal{H}_R\otimes \mathcal{H}_S)$ as 
\begin{equation}
\label{eq:2.4}
\begin{aligned}
\omega_{RS}=\sum_{l=1}^{m} \frac{1}{m} \vert l_{R}\rangle\langle l_{R}\vert\otimes \rho_{S}^{(l)},
\end{aligned}
\end{equation}
where $ \rho_{S}^{(l)}\in \mathcal{S}^{\prime}_{S}$ and $\lbrace \vert l_{R}\rangle\rbrace$ is an orthonormal basis for the ancillary  Hilbert space $\mathcal{H}_{R}$, which we call it the reference Hilbert space.
In addition, the reference state $\omega_{RSE} \in  \mathcal{D}(\mathcal{H}_R\otimes \mathcal{H}_S\otimes\mathcal{H}_E)$ is defined as \cite{7} 
\begin{equation}
\label{eq:2.5}
\begin{aligned}
\omega_{RSE}= \mathrm{id}_R\otimes\Lambda_S (\omega_{RS})
=\sum_{l=1}^{m} \frac{1}{m} \vert l_{R}\rangle\langle l_{R}\vert\otimes \rho_{SE}^{(l)},
\end{aligned}
\end{equation}
where $\mathrm{id}_R$ is the identity map on $\mathcal{L}(\mathcal{H}_R)$,  and $ \rho_{SE}^{(l)}\in \mathcal{S}^{\prime}$ is such that $ \mathrm{Tr}_{E}( \rho_{SE}^{(l)})=\rho_{S}^{(l)}$.

An immediate consequence of the above definitions is that we can construct subspaces $\mathcal{V}_S$ and 
$\mathcal{V}$ as 
 the\textit{ generalized steered sets}, from $\omega_{RS}$ and $\omega_{RSE}$, respectively. We have \cite{7}
\begin{equation}
\label{eq:2.6}
\begin{aligned}
\mathcal{V}_{S}=\left\lbrace \mathrm{Tr}_{R}[(A_{R}\otimes I_{S})\omega_{RS}] \right\rbrace ,  
\end{aligned}
\end{equation}
and
\begin{equation}
\label{eq:2.7}
\begin{aligned}
\mathcal{V}=\left\lbrace \mathrm{Tr}_{R}[(A_{R}\otimes I_{SE})\omega_{RSE}] \right\rbrace ,  
\end{aligned}
\end{equation}
where $ A_{R}$ are arbitrary linear operators in $\mathcal{L}(\mathcal{H}_{R})$, and $I_S$ and $I_{SE}$ are the identity operators on $\mathcal{H}_{S}$ and $\mathcal{H}_{S}\otimes\mathcal{H}_{E}$, respectively.

As we have seen in the previous subsection, when $m< {(d_S)}^2$, we can choose $\rho_{SE}^{(i)}$, 
$i=m+1, \ldots , (d_S)^2$,  arbitrarily. So, there are infinitely many different possible Hermitian assignment maps $\Lambda_S: \mathcal{L}(\mathcal{H}_S)\rightarrow\mathcal{L}(\mathcal{H}_S\otimes\mathcal{H}_E)$ such that their action on $\mathcal{V}_S$ are the same, but they  act differently on (some) operators $y\notin \mathcal{V}_S$.
Therefore, it is possible that we choose a non-CP assignment map $\Lambda_S$, while there exists another 
 assignment map $\Lambda_S^{(CP)}$, which is CP.
In the next subsection, using the reference state $\omega_{RSE}$,  we will see how we can avoid such inappropriate choosing of  the assignment map $\Lambda_S$.

Note that one can use the Choi matrix (operator) \cite{19bb} to determine whether a given linear map, e.g., the assignment map $\Lambda_S$, in our case, is CP or not. 
Construct the operator 
\begin{equation}
\label{eq:2.8a}
\Omega_{RSE}=\mathrm{id}_R\otimes\Lambda_S(\vert\xi_{RS}\rangle\langle\xi_{RS}\vert),
\end{equation}
where  
 the ket $\vert\xi_{RS}\rangle=\frac{1}{\sqrt{d_S}} \sum_{l=1}^{d_S}  \vert l_R\rangle\vert l_S\rangle \in \cH_R\otimes\cH_S$   is  the \textit{maximally entangled} state \cite{25-a}.  (The set $\lbrace \vert l_{S}\rangle\rbrace$ is an orthonormal basis for $\cH_S$.) 
 The assignment map $\Lambda_S$ is CP if and only if the Choi operator is a density operator, i.e.,  $\Omega_{RSE} \in \mathcal{D}(\cH_R \otimes \cH_S \otimes \cH_E)$.
 
In summary, using the Choi operator $\Omega_{RSE}$, in Eq. \eqref{eq:2.8a}, we can verify whether a given assignment map $\Lambda_S$ is CP or not, while, as we will see in next subsection,  using the reference state $\omega_{RSE}$, in Eq. \eqref{eq:2.5}, we can determine whether there exists a CP assignment map $\Lambda_S^{(CP)}$ or not.

\subsection{Markovianity of the reference state and CP-ness of the reduced dynamics}

A tripartite state $\sigma_{RSE}\in  \mathcal{D}(\mathcal{H}_R\otimes \mathcal{H}_S\otimes\mathcal{H}_E)$ is  called a \textit{Markov state}  if  there exists a decomposition of  $\mathcal{H}_{S}$ as $\mathcal{H}_{S}=\bigoplus_{k}\mathcal{H}_{S_{k}}=\bigoplus_{k}\mathcal{H}_{s^{L}_{k}}\otimes\mathcal{H}_{s^{R}_{k}}$  such that
\begin{equation}
\label{eq:2.8}
\sigma_{RSE}=\bigoplus_{k}\lambda_{k}\:\sigma_{Rs^{L}_{k}}\otimes\sigma_{s^{R}_{k}E},
\end{equation}
where $\lbrace \lambda_{k}\rbrace$ is a probability distribution ($\lambda_{k}\geq 0$, $\sum_{k}\lambda_{k}=1$), $\sigma_{Rs^{L}_{k}}\in \mathcal{D} (\mathcal{H}_{R}\otimes\mathcal{H}_{s^{L}_{k}})$ and $\sigma_{s^{R}_{k}E} \in \mathcal{D}(\mathcal{H}_{s^{R}_{k}}\otimes\mathcal{H}_{E})$ \cite{8}. 
In other words, the state  $\sigma_{RSE}$ is a  convex combination of density operators $\sigma_{RS_{k}E}=\sigma_{Rs^{L}_{k}}\otimes\sigma_{s^{R}_{k}E}$, each on a distinct Hilbert space $\mathcal{H}_R\otimes \mathcal{H}_{S_k}\otimes\mathcal{H}_E$.

A Markov state  $\sigma_{RSE}$ can be written as \cite{8}
\begin{equation}
\label{eq:2.9}
\begin{aligned}
\sigma_{RSE}= \mathrm{id}_R\otimes\Lambda_S^{(CP)} (\sigma_{RS}),
\end{aligned}
\end{equation}
where $\sigma_{RS}=\mathrm{Tr}_{E}(\sigma_{RSE})$, and 
 $\Lambda_S^{(CP)}: \mathcal{L}(\mathcal{H}_S)\rightarrow\mathcal{L}(\mathcal{H}_S\otimes\mathcal{H}_E)$
 is a CP (assignment) map. The explicit form of  $\Lambda_S^{(CP)}$ is as
\begin{equation}
\label{eq:2.10}
\begin{aligned}
\Lambda_S^{(CP)}=\bigoplus_{k}\mathrm{id}_{s_{k}^{L}}\otimes \Lambda_{s_{k}^{R}}^{(CP)},
\end{aligned}
\end{equation}
where $\mathrm{id}_{s_{k}^{L}}$ is the identity map on $\mathcal{L}(\mathcal{H}_{ s_{k}^{L}})$, and $\Lambda_{s_{k}^{R}}^{(CP)}: \mathcal{L}(\mathcal{H}_{s_{k}^{R}})\rightarrow  \mathcal{L}(\mathcal{H}_{s_{k}^{R}}\otimes \mathcal{H}_{E})$ is a CP map 
such that $\Lambda_{s_{k}^{R}}^{(CP)}(\sigma_{s^{R}_{k}})=\sigma_{s^{R}_{k}E}$, where $\sigma_{s^{R}_{k}}=\mathrm{Tr}_{E}(\sigma_{s^{R}_{k}E})$ \cite{9-a}.

Comparing Eqs. \eqref{eq:2.5} and \eqref{eq:2.9} shows that if the reference state $\omega_{RSE}$ is a Markov state, then there exists a CP assignment map $\Lambda_{S}^{(CP)}$ such that $\Lambda_{S}^{(CP)}( \rho_{S}^{(l)})= \rho_{SE}^{(l)}$, for all $ \rho_{S}^{(l)}\in \mathcal{S}^{\prime}_S$, i.e., this  $\Lambda_{S}^{(CP)}$ maps $\mathcal{V}_S$ to $\mathcal{V}$. Therefore, the reduced dynamics in Eq. \eqref{eq:1.3} is CP, for arbitrary evolution $U$.

Reversely, if there exists a CP assignment map $\Lambda_{S}^{(CP)}$,  which maps all $ \rho_{S}^{(l)}\in \mathcal{S}^{\prime}_S$ to $\rho_{SE}^{(l)}\in \mathcal{S}^{\prime}$, then the reference state $\omega_{RSE}$, in Eq. \eqref{eq:2.5}, is a Markov state, even if we have used a non-CP assignment map 
 $\Lambda_{S}$ to construct it.

In summary \cite{7}:
\begin{prop}
\label{pro:2.1}
There exists a CP assignment map  $\Lambda_S^{(CP)}: \mathcal{L}(\mathcal{H}_S)\rightarrow\mathcal{L}(\mathcal{H}_S\otimes\mathcal{H}_E)$, which maps $\mathcal{V}_S$ to $\mathcal{V}$ if and only if the reference state $\omega_{RSE}$, in Eq. \eqref{eq:2.5}, is a Markov state, as Eq. \eqref{eq:2.8}.
\end{prop}

Therefore, by checking whether  $\omega_{RSE}$, in Eq. \eqref{eq:2.5}, can be written as  Eq. \eqref{eq:2.8} or not, we can deduce whether there exists a CP assignment map, which maps $\mathcal{V}_S$ to $\mathcal{V}$, or not.

When $\omega_{RSE}$ is a Markov state, and so, there exists a CP assignment map $\Lambda_S^{(CP)}$, then, since $\mathrm{Tr}_{E}$ and $\mathrm{Ad}_{U}$ are also CP \cite{3},  the reduced dynamics of the system  $\mathcal{E}_S$, in Eq. \eqref{eq:1.3},  for any $\rho_S \in  \mathrm{Tr}_E( \mathcal{D}(\mathcal{H}_S\otimes \mathcal{H}_E) \cap \mathcal{V})= \mathcal{D}(\mathcal{H}_S)\cap \mathcal{V}_S$, is CP,  for arbitrary $U$.

In fact, based on the Proposition \ref{pro:2.1}, one can prove that the reduced dynamics of the system $\mathcal{E}_S$, in Eq. \eqref{eq:1.3},  for any $\rho_S \in  \mathrm{Tr}_E( \mathcal{D}(\mathcal{H}_S\otimes \mathcal{H}_E) \cap \mathcal{V})$, is CP, for arbitrary $U$, if and only if $\omega_{RSE}$, in Eq. \eqref{eq:2.5}, is a Markov state \cite{10, 11,  7}.
This result includes all the previous results, on the CP-ness of the reduced dynamics, given in \cite{24, 12cc, 14cc, 13cc}, as special cases \cite{11, 17}.

Finally, note that $\Lambda_S^{(CP)}$, in Eq. \eqref{eq:2.10}, though is CP on the whole $\mathcal{L}(\mathcal{H}_S)$, but is not consistent on the whole $\mathcal{L}(\mathcal{H}_S)$. This is so because 
$\mathrm{id}_{s_{k}^{L}}$,  in Eq. \eqref{eq:2.10}, is, in fact, the projector onto the subspace $\mathcal{L}(\mathcal{H}_{ s_{k}^{L}})$. 
$\Lambda_{s_{k}^{R}}^{(CP)}$, also, includes a projection onto the subspace $\mathcal{L}(\mathcal{H}_{s_{k}^{R}})$.

 $\Lambda_S^{(CP)}$, in Eq. \eqref{eq:2.10}, is consistent on the whole $\mathcal{L}(\mathcal{H}_S)$ only when the summation in  Eq. \eqref{eq:2.10} includes only one term, i.e.,  
$\mathcal{H}_{S}=\mathcal{H}_{s^{L}}\otimes\mathcal{H}_{s^{R}}$, where $\mathcal{H}_{s^{R}}$ is a trivial one dimensional Hilbert space. Then, $\Lambda_S^{(CP)}$, in Eq. \eqref{eq:2.10}, becomes
\begin{equation}
\label{eq:2.11}
\begin{aligned}
\Lambda_S^{(CP)}=\mathrm{id}_{S}\otimes \sigma_{E},
\end{aligned}
\end{equation}
where $\mathrm{id}_{S}$ is the identity map on the whole $\mathcal{L}(\mathcal{H}_S)$, and $\sigma_{E}\in \mathcal{D}(\mathcal{H}_E)$ is a (fixed) state. The above assignment map is the same as the Pechukas's one, in Eq. \eqref{eq:1.4}.

  In summary, using  the fact that any CP assignment map $\Lambda_S^{(CP)}$ can be written as Eq. \eqref{eq:2.10} \cite{11-a}, we have:
\begin{cor}
\label{cor:2.2}
If the assignment map $\Lambda_S: \mathcal{L}(\mathcal{H}_S)\rightarrow\mathcal{L}(\mathcal{H}_S\otimes\mathcal{H}_E)$ is CP and consistent on the whole $\mathcal{L}(\mathcal{H}_S)$, then it is given by Eq. \eqref{eq:2.11}, which is the Pechukas's one.
\end {cor}

But, in proving the  Pechukas's theorem, in \cite{1, 6}, (in addition to the consistency on the whole $\mathcal{L}(\mathcal{H}_S)$) only the positivity of the $\Lambda_{S}$ is assumed. 
In the next section, we will show how existence of a positive assignment map $\Lambda_S^{(P)}$ leads to 
existence of a CP assignment map $\Lambda_S^{(CP)}$.

Let us end this section with the following remark:
\begin{rem}
Our discussions in this section can be, readily, generalized to the case that the environment is infinite dimensional. We have only used this fact that both $\mathcal{V}$ and $\mathcal{V}_S$ have the same finite dimension $m$. In addition, in generalization of $\Lambda_S$ to the whole $\mathcal{L}(\mathcal{H}_S)$, we have only used this fact that the system $S$ is $d_S$-dimensional. 
\end{rem}

\section{Main result}\label{sec: C}
 
\subsection{Markov states and strong subadditivity}

An  important relation in quantum information theory is the \textit{strong subadditivity relation},  i.e., for any tripartite quantum state
 $\sigma_{RSE} \in  \mathcal{D}(\mathcal{H}_R\otimes \mathcal{H}_S\otimes\mathcal{H}_E)$, the following inequality holds \cite{12-1, 3}:
\begin{equation}
\label{eq:3.1}
\begin{aligned}
S(\sigma_{RS})+ S(\sigma_{SE})-S(\sigma_{RSE})-S(\sigma_{S}) \geq 0, 
\end{aligned}
\end{equation}
where $\sigma_{RS}=\mathrm{Tr}_{E}(\sigma_{RSE})$, $\sigma_{SE}=\mathrm{Tr}_{R}(\sigma_{RSE})$  
and $\sigma_{S}=\mathrm{Tr}_{RE}(\sigma_{RSE})$ are the reduced states
 and $S(\sigma)\equiv - \mathrm{Tr}(\sigma \mathrm{log}_2 \sigma)$ is the von Neumann entropy \cite{3}.

The  \textit{relative entropy} of the state $\rho$ to another state $\sigma$ is defined as \cite{3}
\begin{equation}
\label{eq:3.2}
\begin{aligned}
S(\rho\vert\vert\sigma)= \mathrm{Tr}(\rho\mathrm{log}_2\rho)-\mathrm{Tr}(\rho\mathrm{log}_2\sigma),
\end{aligned}
\end{equation}
if $\mathrm{supp}[\rho]\subseteq \mathrm{supp}[\sigma]$, otherwise it is defined to be $+\infty$. 
  ($\mathrm{supp}[\tau]$, the support of  the state $\tau \in  \mathcal{D}(\mathcal{H})$, is the subspace of $\mathcal{H}$, spanned by the eigenvectors of 
$\tau$ with nonzero eigenvalues.)

Using Eq. \eqref{eq:3.2}, it can be shown that $S(\sigma_{RS}\vert\vert\sigma_R\otimes\sigma_{S})=S(\sigma_R)+ S(\sigma_S)- S(\sigma_{RS})$, where  $\sigma_{R}=\mathrm{Tr}_{S}(\sigma_{RS})$. In addition, $S(\sigma_{RSE}\vert\vert\sigma_R\otimes\sigma_{SE})=S(\sigma_R)+ S(\sigma_{SE})- S(\sigma_{RSE})$.
So,  Eq. \eqref{eq:3.1} can be rewritten as
\begin{equation}
\label{eq:3.3}
\begin{aligned}
S(\sigma_{RSE}\vert\vert\sigma_R\otimes\sigma_{SE}) \geq S(\sigma_{RS}\vert\vert\sigma_R\otimes\sigma_{S}).
\end{aligned}
\end{equation}

In Ref. \cite{8}, it has been shown that the strong subadditivity relation,  Eq. \eqref{eq:3.1} or equivalently  Eq. \eqref{eq:3.3}, holds with equality if and only if  $\sigma_{RSE}$ is a Markov state, as Eq. \eqref{eq:2.8}.

Each  tripartite state  $\sigma_{RSE}$  satisfies Eq.  \eqref{eq:3.3}. So, if, in addition, we have 
\begin{equation}
\label{eq:3.4}
\begin{aligned}
S(\sigma_{RSE}\vert\vert\sigma_R\otimes\sigma_{SE}) \leq S(\sigma_{RS}\vert\vert\sigma_R\otimes\sigma_{S}),
\end{aligned}
\end{equation}
then $\sigma_{RSE}$ is a Markov state.
In the next subsection, we will examine the Markovianity of the reference state $\omega_{RSE}$, in Eq. \eqref{eq:2.5}, using  Eq. \eqref{eq:3.4}.

\subsection{Reference state and strong subadditivity} \label{sec:C.2}

For the reference state $\omega_{RSE}$, in Eq. \eqref{eq:2.5}, we have $\omega_{R}=\frac{I_R}{m}$, where $I_R$ is the identity operator on $\mathcal{H}_{R}$. In addition, $\omega_{SE}=\sum_{l=1}^{m} \frac{1}{m}  \rho_{SE}^{(l)}$  and  $\omega_{S}=\sum_{l=1}^{m} \frac{1}{m}  \rho_{S}^{(l)}$. So, $\omega_{SE}=\Lambda_S(\omega_{S})$. Therefore, for $\omega_{RSE}$ in Eq. \eqref{eq:2.5},   Eq. \eqref{eq:3.4} can be rewritten as
\begin{equation}
\label{eq:3.5}
\begin{aligned}
S(\omega_{RSE}\vert\vert\frac{I_R}{m}\otimes\Lambda_S(\omega_{S})) \leq S(\omega_{RS}\vert\vert\frac{I_R}{m}\otimes\omega_{S}).
\end{aligned}
\end{equation}

Now, using Eq. \eqref{eq:2.4}, it can be shown that
\begin{equation}
\label{eq:3.6}
\begin{aligned}
 S(\omega_{RS}\vert\vert\frac{I_R}{m}\otimes\omega_{S})=  S(\sum_{l=1}^{m} \frac{1}{m} \vert l_{R}\rangle\langle l_{R}\vert\otimes \rho_{S}^{(l)}\vert\vert\frac{I_R}{m}\otimes\omega_{S})
 \\ =\sum_{l=1}^{m} \frac{1}{m} S(\rho_{S}^{(l)}\vert\vert \omega_{S}). \qquad\qquad\quad\quad
 \end{aligned}
\end{equation}
Similarly, using Eq. \eqref{eq:2.5}, we have
\begin{equation}
\label{eq:3.7}
\begin{aligned}
 S(\omega_{RSE}\vert\vert\frac{I_R}{m}\otimes\omega_{SE})=  S(\sum_{l=1}^{m} \frac{1}{m} \vert l_{R}\rangle\langle l_{R}\vert\otimes \rho_{SE}^{(l)}\vert\vert\frac{I_R}{m}\otimes\omega_{SE})
 \\ =\sum_{l=1}^{m} \frac{1}{m} S(\rho_{SE}^{(l)}\vert\vert \omega_{SE})  \quad\qquad\qquad\quad \\
 = \sum_{l=1}^{m} \frac{1}{m} S(\Lambda_S(\rho_{S}^{(l)})\vert\vert \Lambda_S(\omega_{S})). \quad\quad
 \end{aligned}
\end{equation}

In Ref. \cite{12}, it has been shown that the relative entropy is monotone, not only under CP maps, but also under positive maps. So, if there exists a positive assignment map  $\Lambda_S^{(P)}: \mathcal{L}(\mathcal{H}_S)\rightarrow\mathcal{L}(\mathcal{H}_S\otimes\mathcal{H}_E)$, which maps $\mathcal{V}_S$ to $\mathcal{V}$, then we have 
\begin{equation}
\label{eq:3.8}
\begin{aligned}
S(\rho_{S}^{(l)}\vert\vert \omega_{S})\geq S(\Lambda_S^{(P)}(\rho_{S}^{(l)})\vert\vert \Lambda_S^{(P)}(\omega_{S})) = S(\rho_{SE}^{(l)}\vert\vert \omega_{SE}).
\end{aligned}
\end{equation}

So, using Eqs. \eqref{eq:3.6}, \eqref{eq:3.7}, and \eqref{eq:3.8}, we achieve Eq. \eqref{eq:3.5}.
Therefore the reference state $\omega_{RSE}$, in Eq. \eqref{eq:2.5}, is a Markov state. Now, Proposition      \ref{pro:2.1} states that there exists a CP assignment map $\Lambda_S^{(CP)}: \mathcal{L}(\mathcal{H}_S)\rightarrow\mathcal{L}(\mathcal{H}_S\otimes\mathcal{H}_E)$, which maps $\mathcal{V}_S$ to $\mathcal{V}$.
Therefore, for any $\rho_S \in  \mathrm{Tr}_E( \mathcal{D}(\mathcal{H}_S\otimes \mathcal{H}_E) \cap \mathcal{V})=\mathcal{D}(\mathcal{H}_S) \cap \mathcal{V}_S$, the reduced dynamics of the system, in Eq. \eqref{eq:1.3}, is CP, for arbitrary system-environment evolution $U$. 
In summary, we have proved the following theorem, as our main result in this paper:
\begin{thm}
\label{thm:3.1}
If there exists a linear trace-preserving positive assignment map  $\Lambda_S^{(P)}: \mathcal{L}(\mathcal{H}_S)\rightarrow\mathcal{L}(\mathcal{H}_S\otimes\mathcal{H}_E)$, which maps $\mathcal{V}_S$ to $\mathcal{V}$, then the reference state $\omega_{RSE}$, in Eq. \eqref{eq:2.5}, is a Markov state, as Eq. \eqref{eq:2.8}. So, there exists a linear trace-preserving completely positive (CP) assignment map  $\Lambda_S^{(CP)}: \mathcal{L}(\mathcal{H}_S)\rightarrow\mathcal{L}(\mathcal{H}_S\otimes\mathcal{H}_E)$, which  maps $\mathcal{V}_S$ to $\mathcal{V}$. This results in the CP-ness of the reduced dynamics in Eq. \eqref{eq:1.3}, for any $\rho_S \in  \mathrm{Tr}_E( \mathcal{D}(\mathcal{H}_S\otimes \mathcal{H}_E) \cap \mathcal{V})=\mathcal{D}(\mathcal{H}_S) \cap \mathcal{V}_S$, and  arbitrary system-environment unitary evolution $U$.
\end{thm}

\subsection{Pechukas's theorem}

As stated in Sec.  \ref{subsec: B.2}, when $m< {(d_S)}^2$, there are infinitely many different Hermitian assignment maps  $\Lambda_S: \mathcal{L}(\mathcal{H}_S)\rightarrow\mathcal{L}(\mathcal{H}_S\otimes\mathcal{H}_E)$, which  map $\mathcal{V}_S$ to $\mathcal{V}$.
But, when $m= {(d_S)}^2$, i.e.,  $\mathcal{V}_S=\mathcal{L}(\mathcal{H}_S)$, then there is only one way to construct the assignment map $\Lambda_S$. If we require that this unique assignment map $\Lambda_S$ is positive, then Theorem \ref{thm:3.1} states that $\Lambda_S$ is, in addition, CP. Now, since $\Lambda_S$ is  consistent on the whole  $\mathcal{V}_S=\mathcal{L}(\mathcal{H}_S)$, using Corollary \ref{cor:2.2}, we conclude that $\Lambda_S$ is as Eq. \eqref{eq:2.11}, which is the Pechukas's one.
%
In summary:
\begin{cor}
If the assignment map $\Lambda_S: \mathcal{L}(\mathcal{H}_S)\rightarrow\mathcal{L}(\mathcal{H}_S\otimes\mathcal{H}_E)$ is (a) positive, and (b) consistent on the whole $\mathcal{L}(\mathcal{H}_S)$, then it is given by Eq. \eqref{eq:2.11}.
\end{cor}

\section{Generalization to arbitrary $\mathcal{V}$}  \label{sec: D}

Consider the set $\mathcal{S}=\lbrace\rho_{SE}\rbrace$ of possible initial states of the system-environment. Let us denote the set of linearly independent members of $\mathcal{S}$ by 
$\mathcal{S}^\prime  = \lbrace\rho_{SE}^{(1)}, \rho_{SE}^{(2)}, \, \ldots\ ,  \rho_{SE}^{(M)}\rbrace$, 
where the integer $M$ is $0< M\leq {(d_S)}^2{(d_E)}^2$. Again, the subspace $\mathcal{V}$ is defined as Eq.  \eqref{eq:2.1}. So, for each  $V\in \mathcal{V}$, we have  $V= \sum_{i=1}^M c_i \rho_{SE}^{(i)}$, with complex coefficients $c_i$.

Without loss of generality, we can assume that $\rho_{S}^{(i)}=\mathrm{Tr}_E (\rho_{SE}^{(i)})$, $i=1, \ldots, m$, are also linearly independent, where $\rho_{SE}^{(i)}\in \mathcal{S}^\prime$, and the integer $m$, $0< m\leq {(d_S)}^2$,  is less than $M$. So, the subspace $\mathcal{V}_S=\mathrm{Tr}_E\mathcal{V}$ is spanned by  $\mathcal{S}^\prime_S  = \lbrace\rho_{S}^{(1)}, \ldots\ ,  \rho_{S}^{(m)}\rbrace$.

As before, we define the Hermitian assignment map  $\Lambda_S$ as $\Lambda_S(\rho_{S}^{(i)})=\rho_{SE}^{(i)}$, $i=1, \ldots, m$. This leads to Eq. \eqref{eq:2.2}, i.e., the assignment map  $\Lambda_S$  maps 
 $\mathcal{V}_S$ to the subspace $\hat{\mathcal{V}}\subset \mathcal{V}$, which is spanned by 
 $\lbrace\rho_{SE}^{(1)}, \ldots\ ,  \rho_{SE}^{(m)}\rbrace$.

Note that 
\begin{equation}
\label{eq:4.1}
\begin{aligned}
\mathcal{V}=\hat{\mathcal{V}}\oplus\mathcal{V}_{0},
\end{aligned}
\end{equation}
where, for each $W\in \mathcal{V}_{0}$, we have $\mathrm{Tr}_{E}(W)=0$. So, the most general possible assignment map is as
\begin{equation}
\label{eq:4.2}
\begin{aligned}
\tilde{\Lambda}_{S}=\Lambda_{S}+\mathcal{V}_{0},
\end{aligned}
\end{equation}
where  $\mathcal{V}_{0}$ denotes   arbitrary  $W\in \mathcal{V}_{0}$. 

In addition, if we define the reference states $\omega_{RS}$ and $\omega_{RSE}$ as Eqs. \eqref{eq:2.4} and \eqref{eq:2.5}, respectively, then, as before,  $\mathcal{V}_S$ is given as the generalized steered set from $\omega_{RS}$, i.e.,  Eq. \eqref{eq:2.6}; but, Eq. \eqref{eq:2.7} gives $\hat{\mathcal{V}}$.

Assume that for each $\rho_{SE} \in \mathcal{S}$, the reduced dynamics of the system is given by
a  map $\Psi_S$.  So, for each $\rho_{S}=\mathrm{Tr}_{E}(\rho_{SE})\in \mathcal{S}_S=\mathrm{Tr}_{E}\mathcal{S}$, we have: 
\begin{equation}
\label{eq:4.3}
\begin{aligned}
\rho_{S}^{\prime}=\Psi_S(\rho_{S})=\mathrm{Tr}_{E} \circ \mathrm{Ad}_{U}(\rho_{SE}).  
\end{aligned}
\end{equation}
The obvious requirement that such a map $\Psi_S$ can be defined, is the \textit{$U$-consistency} of the $\mathcal{S}$ \cite{4}, i.e., if for two states $\rho_{SE}, \, \sigma_{SE}\in\mathcal{S}$, we have $\mathrm{Tr}_{E}(\rho_{SE})=\mathrm{Tr}_{E}(\sigma_{SE})=\rho_{S}$, then we must have $\mathrm{Tr}_{E}\circ\mathrm{Ad}_{U}(\rho_{SE})=\mathrm{Tr}_{E}\circ\mathrm{Ad}_{U}(\sigma_{SE})=\Psi_S(\rho_{S})$ \cite{12-a}.

Let us consider the $U$-consistency condition on the whole $\mathcal{V}$, instead of only on $\mathcal{S}$ \cite{12-b}. 
 In Ref. \cite{4}, it has been shown that $\mathcal{V}$ is $U$-consistent, for arbitrary $U$, if and only if  $\mathcal{V}_0=\lbrace 0\rbrace$, i.e., the case studied in Sec. \ref{sec: B.1}.
 But now, where $m<M$, and so $\mathcal{V}_0\neq\lbrace 0\rbrace$, the subspace $\mathcal{V}$ is $U$-consistent, only for a restricted set of unitary operators
  $U\in\mathcal{G} \subset \mathcal{U}(\mathcal{H}_S\otimes\mathcal{H}_E)$,
   where $\mathcal{U}(\mathcal{H}_S\otimes\mathcal{H}_E)$ denotes the set of all unitary operators on $\mathcal{H}_S\otimes\mathcal{H}_E$. When $\mathcal{V}$ is $U$-consistent, for all $U\in \mathcal{G}$, we say that $\mathcal{V}$ is $\mathcal{G}$-consistent \cite{12-c}.

For each $U\in  \mathcal{G}$, the subspace  $\mathcal{V}_0$ is mapped by $\mathrm{Ad}_U$ to $\mathrm{ker} \mathrm{Tr}_E$. So, for each $\rho_S \in  \mathrm{Tr}_E( \mathcal{D}(\mathcal{H}_S\otimes \mathcal{H}_E) \cap \mathcal{V})$, and each $U\in  \mathcal{G}$, using Eq. \eqref{eq:4.2}, the reduced dynamics of the system is given by 
\begin{equation}
\label{eq:4.4}
\begin{aligned}
\rho_{S}^{\prime}=\mathrm{Tr}_{E} \circ \mathrm{Ad}_{U}(\rho_{SE})  \qquad\qquad\qquad\quad\ \\
= \mathrm{Tr}_{E} \circ \mathrm{Ad}_{U} \circ \tilde{\Lambda}_{S} (\rho_{S}) \qquad\qquad\quad\ \ \\
=\mathrm{Tr}_{E} \circ \mathrm{Ad}_{U} \circ \Lambda_{S} (\rho_{S})\equiv\mathcal{E}_S(\rho_{S}),\quad
\end{aligned}
\end{equation}
where $\rho_{SE} \in   \mathcal{D}(\mathcal{H}_S\otimes \mathcal{H}_E) \cap \mathcal{V}$ is such that $\mathrm{Tr}_E(\rho_{SE})=\rho_{S}$. 
Now, since $\mathrm{Tr}_{E}$ and  $\mathrm{Ad}_{U}$ are CP, and $\Lambda_S: \mathcal{L}(\mathcal{H}_S)\rightarrow\mathcal{L}(\mathcal{H}_S\otimes\mathcal{H}_E)$ is Hermitian, the reduced dynamical map $\mathcal{E}_S$ is Hermitian, in general, for each $U\in  \mathcal{G}$.

Finally, if there exists a positive assignment map  $\Lambda_S^{(P)}$, which maps $\mathcal{V}_S$ to $\hat{\mathcal{V}}$, i.e., if $\Lambda_S$ in Eq. \eqref{eq:4.2} is positive, then we can follow the same line of reasoning, as given in Sec. \ref{sec:C.2}, to prove the following theorem, which is the generalization of Theorem \ref{thm:3.1}, to arbitrary $\mathcal{V}$.

\begin{thm}
\label{thm:4.1}
If there exists a linear trace-preserving positive assignment map  $\Lambda_S^{(P)}: \mathcal{L}(\mathcal{H}_S)\rightarrow\mathcal{L}(\mathcal{H}_S\otimes\mathcal{H}_E)$, which maps $\mathcal{V}_S$ to $\hat{\mathcal{V}}$, then the reference state $\omega_{RSE}$, in Eq. \eqref{eq:2.5}, is a Markov state, as Eq. \eqref{eq:2.8}. So, there exists a linear trace-preserving completely positive (CP) assignment map  $\Lambda_S^{(CP)}: \mathcal{L}(\mathcal{H}_S)\rightarrow\mathcal{L}(\mathcal{H}_S\otimes\mathcal{H}_E)$, which  maps $\mathcal{V}_S$ to $\hat{\mathcal{V}}$. This results in the CP-ness of the reduced dynamics in Eq. \eqref{eq:4.4}, for any $\rho_S \in  \mathrm{Tr}_E( \mathcal{D}(\mathcal{H}_S\otimes \mathcal{H}_E) \cap \mathcal{V})=\mathcal{D}(\mathcal{H}_S) \cap \mathcal{V}_S$, and  arbitrary system-environment unitary evolution $U\in \mathcal{G}$.
\end{thm}

\section{Two-level system} \label{sec: E}

As we have seen in Theorems \ref{thm:3.1} and \ref{thm:4.1}, the positivity of the assignment map leads to the Markovianity of the reference state $\omega_{RSE}$, in Eq. \eqref{eq:2.5}. Then, using Eqs. \eqref{eq:2.7} and \eqref{eq:2.8}, we see that each $\rho_{SE} \in \hat{\mathcal{V}}$ can be written as
\begin{equation}
\label{eq:5.1}
\rho_{SE}=\bigoplus_{k} p_{k}\:\rho_{s^{L}_{k}}\otimes\sigma_{s^{R}_{k}E},
\end{equation}
where $\lbrace p_{k}\rbrace$ is a probability distribution, $\rho_{s^{L}_{k}}$ is a state in $\mathcal{D} (\mathcal{H}_{s^{L}_{k}})$ and $\sigma_{s^{R}_{k}E}$ is a \textit{fixed} state  in $\mathcal{D}(\mathcal{H}_{s^{R}_{k}}\otimes\mathcal{H}_{E})$. This result was previously shown in \cite{11}.

In addition, for each $\rho_{SE} \in \mathcal{V}$, using Eqs. \eqref{eq:4.1} and \eqref{eq:5.1}, we have  \cite{17}
\begin{equation}
\label{eq:5.2}
\begin{aligned}
\rho_{SE}=\bigoplus_{k} p_{k}\:\rho_{s^{L}_{k}}\otimes\sigma_{s^{R}_{k}E}+\mathcal{V}_{0},
\end{aligned}
\end{equation}
where  $\mathcal{V}_{0}$ denotes   a  $W\in \mathcal{V}_{0}$ such that $\rho_{SE}$ becomes a valid state in  $\mathcal{D}(\mathcal{H}_{S}\otimes\mathcal{H}_{E})$.

Let us consider the case that the system $S$ is a qubit. So, the three following decompositions of  $\mathcal{H}_{S}$ are possible:

(1) $\mathcal{H}_{S}=\mathcal{H}_{S}\otimes\mathcal{H}_{s^{R}}$, where $\mathcal{H}_{s^{R}}$ is a trivial one dimensional Hilbert space. So,  $\rho_{SE}$, in Eq. \eqref{eq:5.2}, can be written as
\begin{equation}
\label{eq:5.3}
\begin{aligned}
\rho_{SE}=\rho_{S}\otimes\sigma_{E}+\mathcal{V}_{0},
\end{aligned}
\end{equation}
where  $\rho_{S}$ is a state in $\mathcal{D} (\mathcal{H}_S)$ and $\sigma_{E}$ is a fixed state  in $\mathcal{D}(\mathcal{H}_{E})$. Equation \eqref{eq:5.3}, without $\mathcal{V}_{0}$, is the same as Eq. \eqref{eq:1.4}, which is the Pechukas's case \cite{1, 6}.

(2)  $\mathcal{H}_{S}=\mathcal{H}_{s^L}\otimes\mathcal{H}_{S}$, where $\mathcal{H}_{s^{L}}$ is a trivial one dimensional Hilbert space. So,  $\rho_{SE}$, in Eq. \eqref{eq:5.2}, can be written as
\begin{equation}
\label{eq:5.4}
\begin{aligned}
\rho_{SE}=\sigma_{SE}+\mathcal{V}_{0},
\end{aligned}
\end{equation}
where $\sigma_{SE}$ is a fixed state  in $\mathcal{D}(\mathcal{H}_{S}\otimes\mathcal{H}_{E})$.

(3)  $\mathcal{H}_{S}=\mathcal{H}_{S_1}\oplus\mathcal{H}_{S_2}$, where $\mathcal{H}_{S_1}$ and $\mathcal{H}_{S_2}$ are one dimensional Hilbert spaces.
So,  $\rho_{SE}$, in Eq. \eqref{eq:5.2}, can be written as
\begin{equation}
\label{eq:5.5}
\begin{aligned}
\rho_{SE}= p_1\vert 1_{S}\rangle\langle 1_{S}\vert\otimes \sigma_{E}^{(1)}
+  p_2\vert 2_{S}\rangle\langle 2_{S}\vert\otimes \sigma_{E}^{(2)}+\mathcal{V}_{0},
\end{aligned}
\end{equation}
where $\lbrace \vert 1_{S}\rangle, \vert 2_{S}\rangle\rbrace$ is a fixed orthonormal  basis for  $\mathcal{H}_{S}$, and $\sigma_{E}^{(1)}$ and $\sigma_{E}^{(2)}$ are  fixed states  in $\mathcal{D}(\mathcal{H}_{E})$. Equation \eqref{eq:5.5}, without $\mathcal{V}_{0}$, was first introduced in \cite{24}, as a set for which the reduced dynamics is  CP.

Note that in the second case, from  Eq. \eqref{eq:5.4}, we see that there is only one possible initial state for the system as $\rho_{S}=\mathrm{Tr}_{E} (\rho_{SE})=\mathrm{Tr}_{E} (\sigma_{SE})$. In other words, $\mathcal{V}_S$ is one dimensional. So, we neglect this (maybe unimportant) case.

As stated in the previous section, when  $\mathcal{V}_0\neq\lbrace 0\rbrace$, then 
 $\mathcal{G} \neq \mathcal{U}(\mathcal{H}_S\otimes\mathcal{H}_E)$ \cite{4}, i.e., there exists, at least, one $U\in  \mathcal{U}(\mathcal{H}_S\otimes\mathcal{H}_E)$ for which the reduced dynamics of the system  cannot be given by a map.

On the other hand, when  $\mathcal{V}_0 =\lbrace 0\rbrace$, then, from Eqs. \eqref{eq:5.3} and \eqref{eq:5.5}, we see that there is no entanglement \cite{25-a}, between the system $S$ and the environment $E$. Therefore, if 
$\mathcal{D}(\mathcal{H}_S\otimes \mathcal{H}_E) \cap \mathcal{V}$ includes entangled states, then 
  the reference state $\omega_{RSE}$, in Eq. \eqref{eq:2.5}, is not a Markov state, as Eq. \eqref{eq:2.8}.  When the reference state is not a Markov state, then the reduced dynamics of the system  is non-positive, for, at least, one $U$ \cite{25}.

In summary:
\begin{cor}
Consider the case that the system $S$ is a qubit. Neglecting the case that $\mathcal{S}_S$ includes only one $\rho_S$, if $\mathcal{D}(\mathcal{H}_S\otimes \mathcal{H}_E) \cap \mathcal{V}$ includes entangled states, then  either the reduced dynamics of the system  is not given by a map, for, at least, one  $U$, or the reduced dynamics is non-positive, for, at least, one $U$.
\end{cor}

\section{Conclusion} \label{sec: F}

We have considered an arbitrarily chosen (constructed) set $\mathcal{S}=\lbrace\rho_{SE}\rbrace$ of possible initial states of the system-environment.   
Using this $\mathcal{S}$, we have constructed the subspace $\mathcal{V} \subseteq \mathcal{L}(\mathcal{H}_S\otimes\mathcal{H}_E)$, which is spanned by states.
Then, we have seen that for arbitrary unitary time evolution of the system-environment $U\in \mathcal{G}$, the reduced dynamics of the system, for any $\rho_S \in  \mathrm{Tr}_E( \mathcal{D}(\mathcal{H}_S\otimes \mathcal{H}_E) \cap \mathcal{V})$, is given by the map $\mathcal{E}_S=\mathrm{Tr}_{E} \circ \mathrm{Ad}_{U}  \circ \Lambda_S$, which is a Hermitian map, since the assignment map $\Lambda_S$ is Hermitian, in general. 
Note that since $\mathcal{S}_S=\mathrm{Tr}_E\mathcal{S}\subseteq\mathrm{Tr}_E( \mathcal{D}(\mathcal{H}_S\otimes \mathcal{H}_E) \cap \mathcal{V})$, the above result is  valid for our arbitrarily chosen set $\mathcal{S}$, too.

When $\Lambda_S$ is, in addition, CP, then the reduced dynamics is also CP. CP reduced dynamics is commonly used in the quantum information theory \cite{3}, as we have seen in  Introduction,  and in the theory of open quantum systems \cite{14, 15, 16}, e.g., in deriving the GKS-Lindblad master equation \cite{40b, 41b} (where, in fact,  \textit{CP-divisibility} \cite{42b} is used). 
In addition, the CP-ness of $\Lambda_S$ can give us the structures of $\mathcal{V}$,  $\mathcal{V}_S$,  $\mathcal{S}$ and  $\mathcal{S}_S$ 
 \cite{11, 17}.
Therefore, the CP-ness of the assignment map $\Lambda_S: \mathcal{L}(\mathcal{H}_S)\rightarrow\mathcal{L}(\mathcal{H}_S\otimes\mathcal{H}_E)$  is a fruitful result. 

In this paper, using the result of \cite{12}, of  monotonicity of the  relative entropy under positive maps, and using the reference states, in Eqs. \eqref{eq:2.4} and \eqref{eq:2.5}, introduced in \cite{7}, we have shown that the existence of a positive assignment map $\Lambda_S^{(P)}$ results in the  existence of a CP assignment map $\Lambda_S^{(CP)}$.
Therefore, we actually deal with only two types of assignment maps: (a) CP assignment maps, and (b) non-positive Hermitian ones. For a CP assignment map, the reduced dynamics is CP, for any $\rho_S \in  \mathrm{Tr}_E( \mathcal{D}(\mathcal{H}_S\otimes \mathcal{H}_E) \cap \mathcal{V})=\mathcal{D}(\mathcal{H}_S) \cap \mathcal{V}_S$, and all $U\in \mathcal{G}$; but, for a non-positive assignment map it is not necessarily so. In fact, when $\mathcal{G}=\mathcal{U}(\mathcal{H}_S\otimes\mathcal{H}_E)$ and, in addition, the reference state, in Eq. \eqref{eq:2.5}, is not a Markov state, as Eq. \eqref{eq:2.8}, then there exists no CP  assignment map, and the reduced dynamics, for at least one $U$, is non-positive \cite{25, 7}.

As an application, we have considered the case that the system $S$ is a qubit. Neglecting the case that $\mathcal{S}_S$ includes only one $\rho_{S}$, we have shown that when  $\mathcal{V}$ includes entangled states, then
 either the reduced dynamics of the system  is not given by a map, for, at least, one  $U$, or the reduced dynamics is non-positive, for, at least, one $U$.




\begin{thebibliography}{1}
\bibitem{9-a}  For  arbitrary two fixed states, e.g. $\sigma_{s^{R}_{k}}$ and $\sigma_{s^{R}_{k}E}$, one can always find a CP map which maps one to the other \cite{9}.
\bibitem{11-a}  This is a consequence of Proposition \ref{pro:2.1}: CP-ness of $\Lambda_S^{(CP)}$ leads to Markovianity of the reference state, which results in Eq. \eqref{eq:2.10}. This result was previously proven in   \cite{11}. 
Strictly speaking, we should say that the  Markovianity of the reference state results in validity of  the assignment map in Eq.  \eqref{eq:2.10}. It does not exclude the possibility of other (CP) assignment maps, in general. Only when $\mathcal{V}_S =\mathcal{L}(\mathcal{H}_S)$, which is the case used in Corollary \ref{cor:2.2}, the assignment map is unique, and so is as  Eq.  \eqref{eq:2.10}.
\bibitem{12-a}  In Sec. \ref{sec: B.1}, where we have assumed one to one correspondence
  between the members of  $\mathcal{V}$ and $\mathcal{V}_S$, the $U$-consistency condition is trivially satisfied, for arbitrary system-environment unitary evolution $U$. But, in the current section, where, in general, there is no such  one to one correspondence, the   $U$-consistency condition plays an important role.
\bibitem{12-b} It may impose extra restriction on $\mathcal{S}$, unless $\mathcal{S}$ is convex \cite{4}, i.e., there may be a unitary $U$ for which $\mathcal{S}$ is $U$-consistent, while $\mathcal{V}$ is not so.
\bibitem{12-c} Note that, in this paper, we have used the word ``consistent" for two different purposes. In the previous sections, obeying Ref. \cite{2}, we called an assignment map $\Lambda_S$ consistent on  $\mathcal{V}_S$, when for each $x\in\mathcal{V}_S$,
$\Lambda_S(x)=X$ implies $\mathrm{Tr}_E(X)=x$. In the current section, obeying Ref. \cite{4}, we say that a subspace $\mathcal{V}$ is $U$-consistent, when $\mathrm{Ad}_U$ maps  $\mathcal{V}_0$ to $\mathrm{ker} \mathrm{Tr}_E$, the set of all $Z\in \mathcal{L}(\mathcal{H}_{S}\otimes\mathcal{H}_{E})$ for which we have $\mathrm{Tr}_{E}(Z)=0$.
\end{thebibliography}
\end{document}